\documentclass[twocolumn,amsmath]{revtex4-1}
\makeatother
\usepackage[dvips]{graphicx}

\usepackage{amsmath}
\usepackage{amsbsy}
\usepackage{caption}
\usepackage{color}
\usepackage{subcaption}
\DeclareGraphicsExtensions{.pdf,.eps,.png,.jpg,.mps}

\definecolor{textcolor}{cmyk}{0,0,0,1}
\definecolor{magenta}{rgb}{1,0,1}
\definecolor{green}{rgb}{0,1,0}
\definecolor{red}{rgb}{1,0,0}

\begin{document}

\title{
Chemical Bonding of Transition-metal Co$_{13}$ Clusters with Graphene
}
\author{ T. Alonso-Lanza,$^{*,1}$ A. Ayuela,$^{1}$ F. Aguilera-Granja,$^{1,2}$}
\affiliation{
$^*$tomas\_alonso001@ehu.eus
\\
$^1$Centro de F\'{\i}sica de Materiales CFM-MPC CSIC-UPV/EHU, Donostia 
International Physics Center (DIPC), Departamento de F\'{\i}sica de Materiales, Fac. de Qu\'{\i}micas, UPV-EHU, 20018 San Sebasti\'an, Spain
\\
$^2$Instituto de F\'{\i}sica, Universidad Aut\'onoma de San Luis de Potos\'{\i}, 78000 San Luis Potos\'{\i} S.L.P., M\'exico
\\
}


\begin{abstract}
We carried out density functional calculation to study Co$_{13}$ clusters on graphene. We deposit several free isomers in different disposition respect to hexagonal lattice nodes, studying even the $hcp$ $2d$ isomer recently obtained as the most stable one. Surprisingly, Co$_{13}$ clusters bonded to graphene prefer $icosahedron-like$ structures where the low lying isomer is much distorted, because it is linked with more bonds than in previous works. For any isomer the most stable position binds to graphene by the Co atoms that can lose electrons. We find that the charge transfers between graphene and clusters are small enough to conclude that the Co-graphene binding is not ionic-like but chemical. Besides, the same order of stability among the different isomers on doped graphene is well kept. These findings could also be of interest for magnetic clusters on graphenic nanostructures such as ribbons and nanotubes.
\end{abstract}

\maketitle

\section*{Introduction}

Graphene is a material composed of carbon atoms building a bi-dimensional structure with honeycomb shape, already synthesized and with prospect future applications \cite{novoselov2004electric}.  For instance, the properties of graphene are incredibly promising to develop applications in different fields. Graphene displays properties such as high electron mobility, ballistic transport or the quantum Hall effect \cite{neto2009electronic,novoselov2012roadmap,soldano2010production,singh2011graphene}. All these properties and the easiness to be functionalized grant it as a key material in electronic applications such as the graphene-based transistor \cite{graphene-barristor, schwierz2010graphene}. The spin transport over micrometre-scale distances in single graphene layers \cite{tombros2007electronic} suggests that graphene can be used for spintronics devices, such as spin valves \cite{hill2006graphene, cho2007gate}.
Although pristine graphene is not magnetic, it is possible to turn it magnetic by different means as for example growing nanoislands \cite{fernandez2007magnetism}, nanoribbons \cite{son2006half} or dopping it with structural defects \cite{yazyev2007defect}. Intrinsic magnetism in carbon materials has been proposed, and it has been thoroughly studied \cite{prl-andres, carbon-adatoms-graphene, chen2013towards}.
Another way of changing the electronic and magnetic properties of graphene consists on the adsorption of transition metal atoms \cite{nature-Louie} and clusters \cite{andres-zirconio}; the carbon-transition-metal bonding flavour, ionic or "covalent-like", is decisive for those properties. Graphene magnetism can be tuned by the adsorption of transition metal clusters due to the localized magnetic moments they provide \cite{dai2010absorption}. Here, the topic is to focus on the interaction of magnetic cobalt clusters with graphene.

Clusters of 3d transition metal elements are indeed candidates for developing magnetic applications. Magnetic nanoparticles have been proposed as magnetic bits \cite{sun2000monodisperse}; specifically, a cobalt dimer on a hexagonal carbon ring simulating graphene has been proposed as the basic unit for magnetic data storage \cite{xiao2009co}.
Transition metal clusters have been studied during the past years both theoretically \cite{alonso2000electronic, pastor1995magnetic, pastor1989size, guirado2003orbital, guevara1997electronic, aguilera2008comparative, jones2004classical} and experimentally \cite{apsel1996surface, billas1993magnetic}, being cobalt one of the elements on which recent studies were focused \cite{faustino2003, datta2007-Co-clusters}.
The interest on transition metal clusters rises based on the unexpected electronic and magnetic properties, because their small size allows to appear quantum phenomena not seen on bulk. The valence electrons in transition metal clusters bring high complexity to bonds because itinerant d electrons coexist with delocalized sp electrons.

Some general trends in the variation of cluster properties with size have been proposed. The magnetic moment of transition metal clusters, as for example cobalt clusters \cite{aguilera2006magnetic}, increases from the bulk value when reducing cluster size; this reduction is non-monotonic as there are over imposed oscillations because particular sizes favor larger or lower magnetic moments. The cluster magnetization is complex involving a delicate interplay between electronic and geometrical factors. Cluster magnetism depends on the symmetry, the local coordination and the interatomic distances.
For instance, cobalt clusters with fewer coordinated atoms have enhanced magnetic moment due to the narrowing of the d-band \cite{aguilera2006magnetic}. Note that in the case of small clusters, there is a large difference in the coordination number between inner and surface atoms.

Stern-Gerlach experiments allow to determine the average magnetic moment per atom in a magnetic cluster \cite{apsel1996surface, billas1993magnetic}. However, they do not discriminate between the spin and orbital contributions where X-ray magnetic circular dichroism (XMCD) experiments are quite useful \cite{ohresser2001magnetism, lau2002spin, lau2002size}. In any case, experimental results have to overcome the isomerization effects; as various isomers can be present simultaneously in the experimental setup,  the measured data will be the average of moment values.
The structure of free clusters is not easy to be measured by experiments because the characterization techniques used in bulk systems are unsuitable for free clusters. Therefore, calculations play a key role on finding out the minimum energy isomer of a free cluster. Once the geometries of equilibrium have been theoretically found, comparing calculated values of the magnetic moments with the experimental values is an accepted way to check the obtained results.

Studying computationally the adsorption of transition metal atoms \cite{mao2008density, cao2010transition, wehling2011transition} and clusters \cite{JAlonso-paladio, akturk2009n, okazaki2010structures} on graphene and nanoribbons \cite{longo2011ab} is a field fast evolving. The interaction of defective graphene and clusters of transition metal atoms such as platinum,\cite{fampiou2012binding, kim2011carbon}, iron or aluminium\cite{lim2011dft}, ruthenium\cite{liu2012graphene} and gold\cite{logsdail2011interaction} has also been studied computationally. 

There are also many experimental works focusing on the interaction of transition metals with graphene \cite{zan2011metal, santos2011electronic, eelbo2013adatoms, pi2009electronic, vo2010ultrathin, gan2008one}. It has been recently found experimentally that 2D cobalt planar structures grow between graphene layers \cite{de2014patterning}. Specifically, cobalt-graphene interaction has been studied using cobalt atoms \cite{nature-Louie, wehling2010orbitally, Co-graphene-Krakauer}, also as substitutional impurities \cite{PhysRevB.81.125433}. 

Note that much work has been done on the study of the electronic structures of clusters and alloys of transition metals, including Pt and Pd, deposited over pristine and defective graphene have been extensively investigated before \cite{sahoo2014first, qi2015theoretical, sahoo2015using, ramos2015palladium, liu2014structures, ozaydin2014formation, xu2013density, ji2014first, lópez2014palladium}. However, we are here interested in the coupling of the three traditional ferromagnetic with graphene. They are being 3d materials and among those, we choose cobalt ones because they are not only the standard choice in experiments, but the cobalt bond is clearly different from its close neighbours Fe and Ni.
Here we focus on small cobalt clusters with thirteen atoms. From experiments on magnetism \cite{bucher1991magnetic}, it is known that there is more than one isomer for cobalt clusters.
Some years ago \cite{cantera2010theoretical}, the icosahedron was claimed the minimum energy structure in the cobalt cluster of thirteen atoms, so it was used to be deposited on graphene.\cite{Liu2012} However, it has been recently discovered \cite{faustino-star_like} that the minimum energy geometry consists of a bi-planar structure. This agrees with the fact that, for the late 3d and 4d transition metal elements, clusters with thirteen atoms are more stable in the form of a low symmetry biplanar-like structure \cite{chang2004alternative}. We have then to test how several Co$_{13}$ isomers adsorb on graphene.

In this work we study how the order of stability between the different free isomers of Co$_{13}$ clusters changes when bonding to graphene. 
In the next section we present the results divided into several subsections.
We begin by reproducing that a bi-planar structure is the most stable free Co$_{13}$ cluster isomer.
We then found that among all isomers lying on graphene, icosahedron structure is the most stable.
We explain this finding looking at the charge transfer and the bonding between the cobalt clusters and graphene.
We check the influence of the on-site interaction on our results.
Last, we study how free clusters interact with doped graphene, either shifting the Fermi energy or doping with boron or nitrogen atoms. We thus generalize our conclusions for doped graphene as found in experimental setups by the effect of different substrates. The computational details of density functional calculations are given at the end of the paper.

\section*{Results and discussion}

\subsection{Stability of Free Co$_{13}$ Clusters}

\begin{figure}[h]
\centering
	\includegraphics[width=8cm]{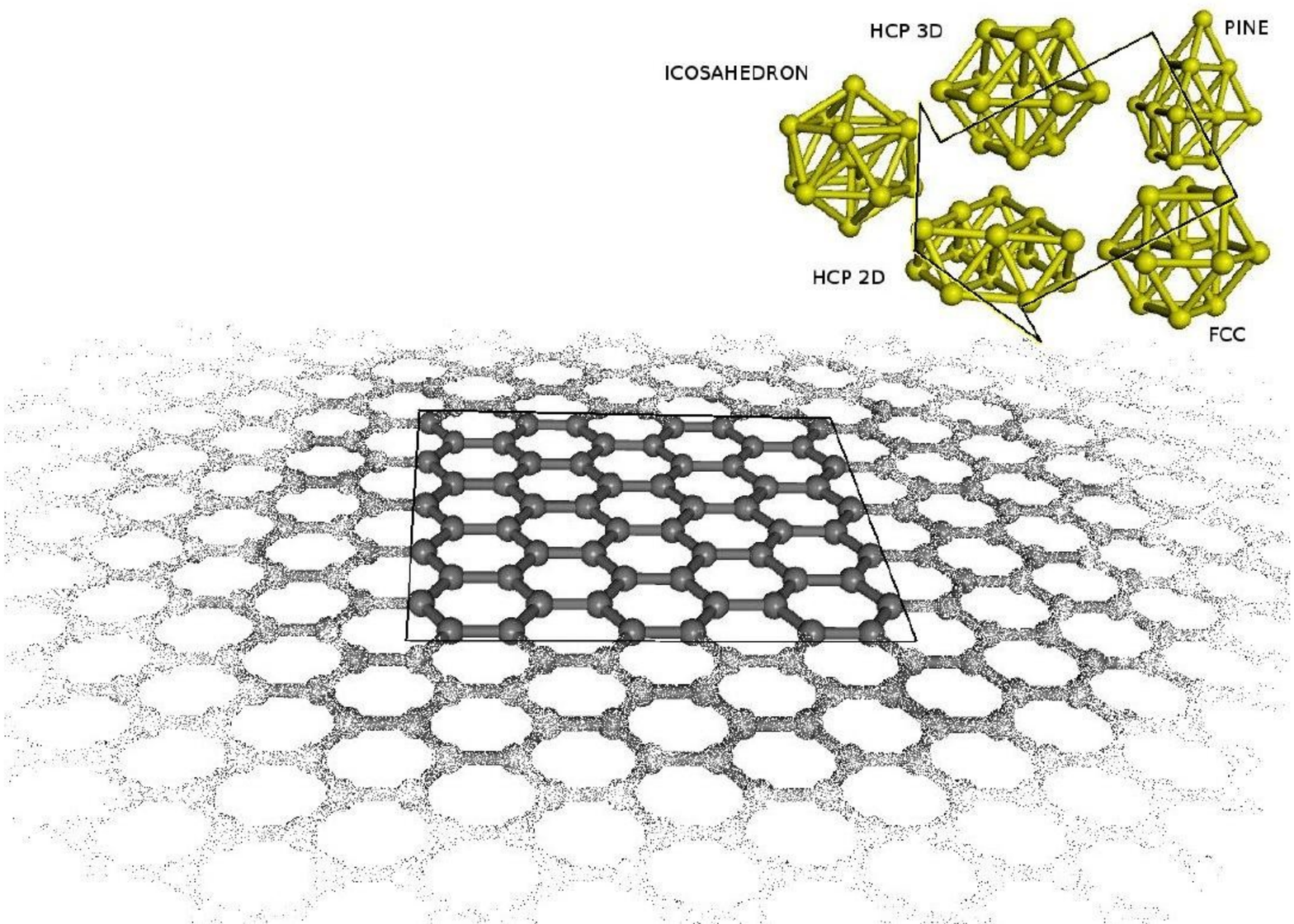}
	\caption{\label{fgr:experimento} 
Image of the proposed experiment: isomers of Co$_{13}$ move toward graphene.} 
   \end{figure}

Experiments evidence isomers for cobalt clusters \cite{bucher1991magnetic}, so we want to deal with the isomerization effects. We have then to recognize the order of stability of the different isomers for the Co$_{13}$ cluster. We already know that for the free Co$_{13}$ cluster, the bi-planar isomer is more stable than the well-known icosahedron \cite{faustino-star_like}.
We have checked this result by testing a wider sample of five different geometries for the free Co$_{13}$ cluster. We have the following order of stability: $hcp$ $2d$, $pine$, $icosahedron$, $fcc$ and $hcp$ $3d$, where the pairs $pine$ - $icosahedron$ and $fcc$ - $hcp$ $3d$ are almost degenerated. These free isomers are displayed in Fig. \ref{fgr:experimento} flying towards graphene, and Table~\ref{tbl:table1} collects their main equilibrium properties. We have obtained that the bi-planar structure, namely $hcp$ $2d$, is the most stable isomer. Note that the so-called $pine$ isomer is equally or even more stable than the $icosahedron$, although nearly degenerate.

\begin{table}[h]
\small
\caption{\ Values of the binding energy ($E_{b_{free}}$), the total magnetic moment ($\mu_{T}$), the average magnetic moment per cobalt atom ($\mu_{av}$), the average bond length ($d_{av}$), and the average coordination number ($coord_{av}$) for the five different free isomers of Co$_{13}$ in decreasing order of stability.}
\label{tbl:table1}
\begin{tabular}{llllll}
\hline
isomer & $E_{b_{free}}$  (eV)  & $\mu_{T} ( \mu_{B})$ & $\mu_{av}$ ( $\mu_{B})$ & $d_{av} (\mathring{A})$  & $coord_{av}$ \\
\hline
$hcp$ $2d$ & 3.508 &  26.954 & 2.073 & 2.44 &  5.54\\
$pine$ & 3.452 & 23.000 & 1.769 & 2.40 &  5.54\\
$icosahedron$ & 3.444 &  30.974 & 2.383 & 2.48 &  6.31\\
$fcc$ & 3.413 &  27.000 & 2.077 & 2.42 &  5.54\\
$hcp$ $3d$ & 3.406 &  25.000 & 1.923 & 2.42 &  5.54\\
\hline
  \end{tabular}
\end{table}

The values of the binding energy for free clusters are similar in magnitude to those of previous calculations \cite{Liu2012}, while the values of average bond length are slightly larger\cite{Liu2012, faustino2003}.
The magnetic moment per atom depends on the specific isomer. We got 2.07 $\mu_{B}$ for the $hcp$ $2d$ and 2.38 $\mu_{B}$ for the $icosahedron$. The icosahedron value is close to that reported previously \cite{datta2007-Co-clusters}. Experimental values were reported to be about 2 $\mu_{B}$ \cite{xu2005magnetic, peredkov2011x}, which are close to the value we have obtained for $hcp$ $2d$, confirming it as the ground state rather than the icosahedra. The large magnetic moment of the $icosahedron$ cluster can be explained as the effect of narrowing the d bands due to larger average bond lengths. We can see in Table~\ref{tbl:table1} that the $icosahedron$ isomer presents the highest coordination and the largest average bond length, due to the Co-Co repulsion between central and surface atoms.
As found in previous publications \cite{chang2004alternative}, the bi-planar structure $hcp$ $2d$ is preferred here over the $icosahedron$ because it allows to reduce both the electron and ion repulsion. The latter term becomes dominant for the late 3d transition metal elements over the term that accounts for the interaction between electrons and ionic cores, which favors the $icosahedron$. Thus, this two terms balance results in the $hcp$ $2d$ isomer being the most stable one for Co$_{13}$-free clusters.

\subsection{
Stability of Co$_{13}$ Isomers Bonded to Graphene}

\begin{figure*}[htpb]
\centering
	\includegraphics[width=13.5cm]{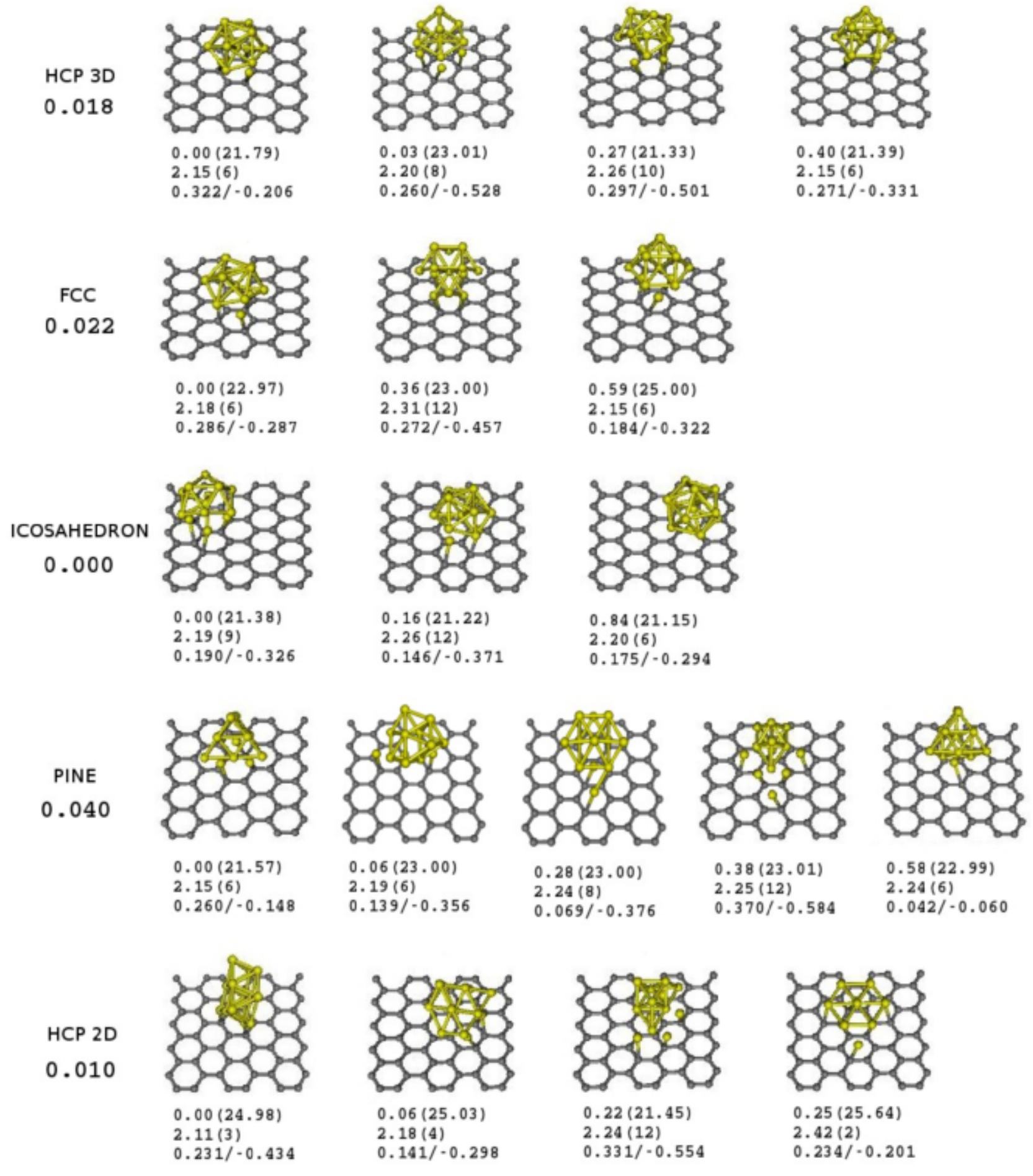}
	\caption{\label{fgr:geometries} 
Relaxed geometries for each isomer of Co$_{13}$. For each of the geometries we show below in the first line the energy difference in eV respect to the lowest energy geometry of each isomer and the total spin polarization; in the second line, the average Co-C bond length in $\mathring{A}$ and the number of Co-C bonds; in the third line, the number of electrons gained by graphene and the induced magnetic moment on graphene. Besides the name of the isomer, it is shown the relative energy in eV of the minimum energy geometry for each isomer respect to the global minimum, which is the icosahedron.}  
	\end{figure*}

\begin{figure*}[htpb]
\centering
	\includegraphics[width=13.5cm]{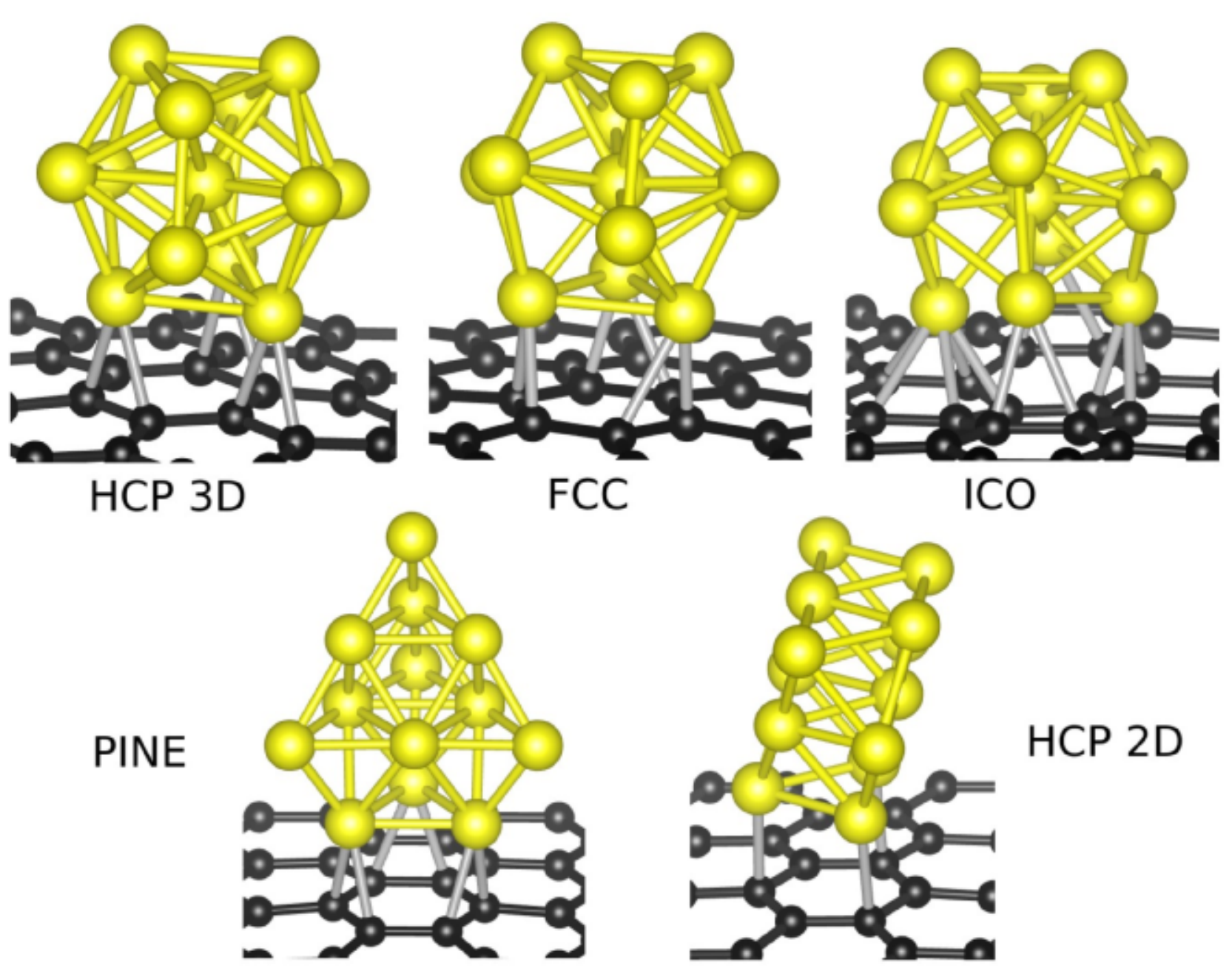}
	\caption{\label{fgr:geometries2} 
For each isomer, the interfacial structure of the most stable arrangement lying over graphene is shown.}  
	\end{figure*}

When the Co$_{13}$ isomers attach to graphene, their geometries are expected to suffer structural changes. Figure \ref{fgr:geometries} shows the relaxed geometries for each isomer of Co$_{13}$ cluster deposited on graphene with several arrangements. The dispositions found more stable for each isomer are enlarged in Fig. \ref{fgr:geometries2}. We have tried all the possible configurations for each isomer. For example, for the icosahedron we place it on graphene by a vertex, an edge or by a triangular face. All these few deposition structures were considered. However, in the case of the pine or the hcp2d, different faces are found, and each of them together with the vertex one has also been computed on graphene. The $hcp$ $2d$ cluster, which is the ground state of the free clusters, falls apart from the low-energy state in three minima: one at 0.06 eV and other two that are nearly degenerate at 0.22 and 0.25 eV. The $icosahedron$ presents a first minima at 0.16 eV with respect to the ground state.
The ground state consists of a relaxed icosahedron-like cluster that loses the high symmetry inherent to the free state, where they are regular icosahedra. The cluster is bonded to graphene by an edge of the icosahedron, meaning that two cobalt atoms directly bond carbon atoms. During relaxation other two cobalt atoms, those forming triangular faces with the two previous atoms, move quite closer to graphene to obtain a largely stable structure with four cobalt atoms bonding graphene. This structure is totally different from other previously reported\cite{Liu2012} with a vertex of the icosahedron placed over a hollow and the bonding reinforced by two near cobalt atoms, which has a total of three cobalt atoms involved directly in the bonding. Of course the previously reported structure appears in our calculations, but it is not the lowest lying icosahedral-like structure.

Other Co$_{13}$ isomers classify in a similar class because they present energy differences between structures following the same patterns. The $pine$ and $fcc$ isomers have structures at 0.38-0.58 eV and at 0.36-0.59 eV, respectively; the $pine$ has other two minima at 0.06 and 0.28 eV. The $hcp$ $3d$ structures seem also to belong to this last pattern with energies between minima of 0.03, 0.27 eV and 0.40 eV. 
If we focus on the most stable geometry over graphene for each isomer, it is worth mentioning that high-energy isomers in free Co$_{13}$ clusters ($fcc$ and $hcp$ $3d$) relaxed their geometries largely towards distorted icosahedra. Note that the most stable $hcp$ $2d$ and $pine$ as free clusters remain almost similar when deposited on graphene.

\begin{figure}[h]
\begin{subfigure}{.5\textwidth}
  \centering
  \includegraphics[width=8cm]{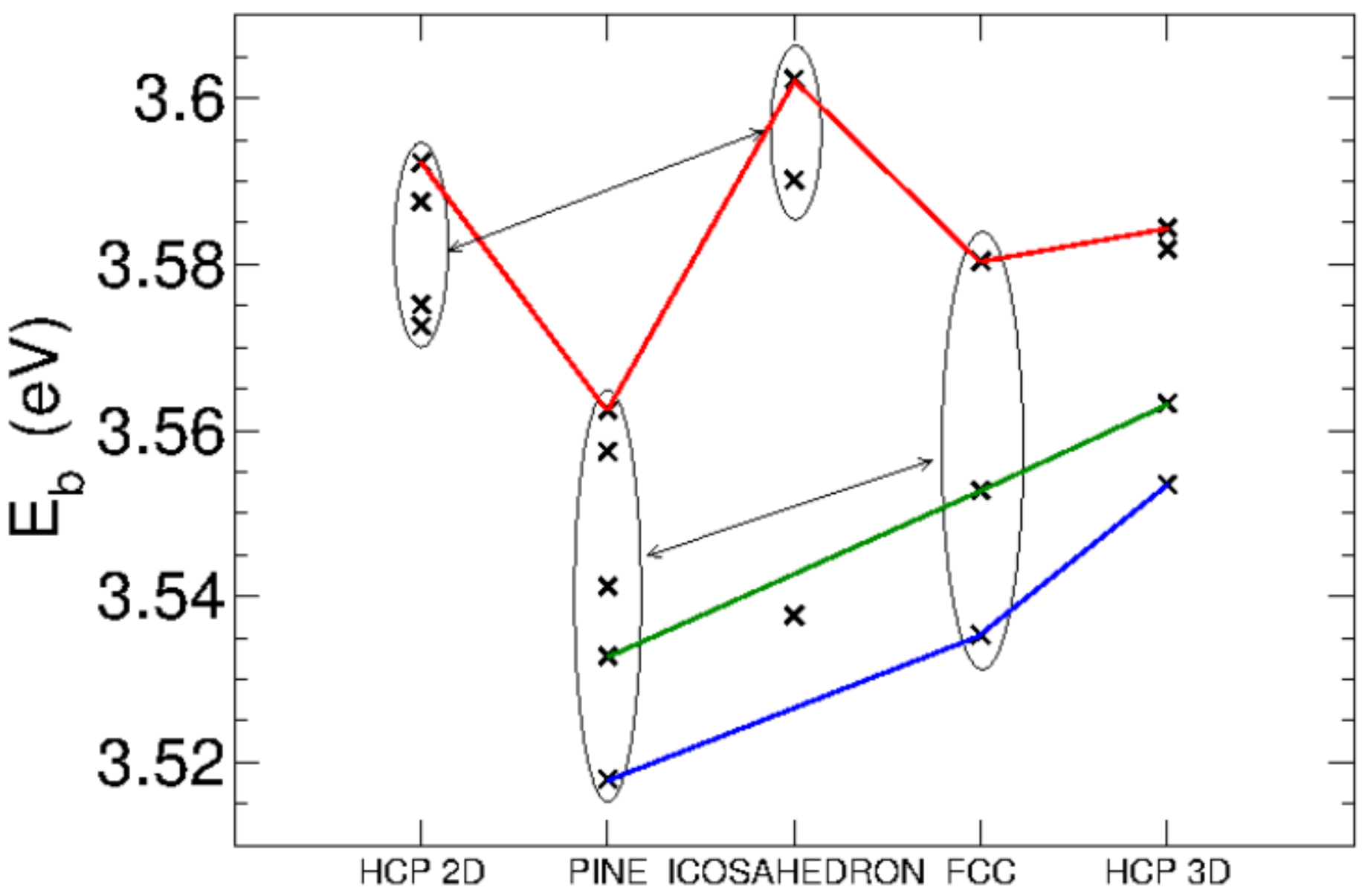}
  \caption{\label{fgr:binding}}  
\end{subfigure}\\
\begin{subfigure}{.5\textwidth}
  \centering
  \includegraphics[width=8cm]{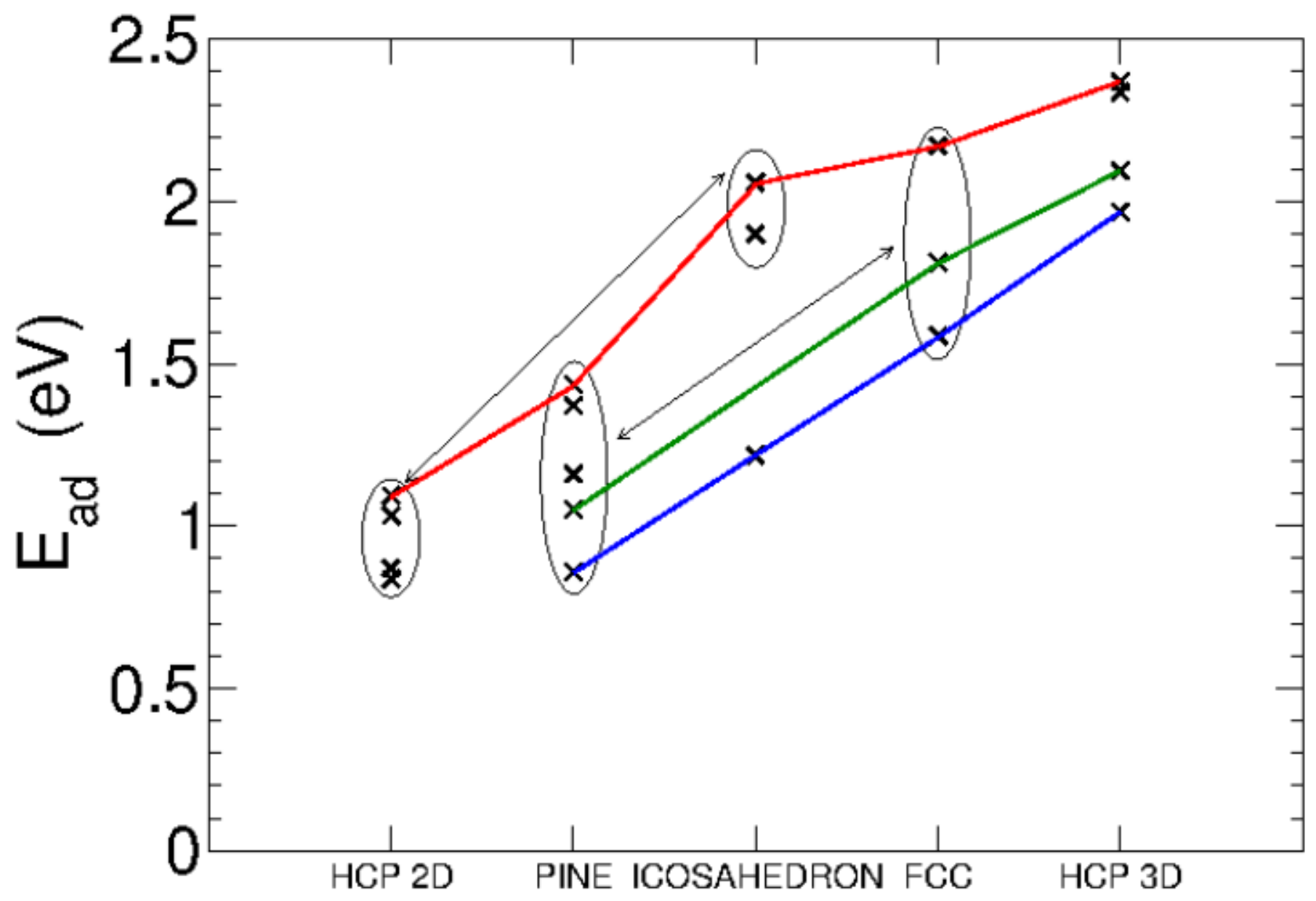}
  \caption{\label{fgr:adsorption}} 
\end{subfigure}
\caption{\label{fig:fig3}
(Colour online) 
(a) Binding energy of the different calculated geometries for each isomer of  Co$_{13}$ on graphene. (b) Adsorption energies of the different calculated geometries for each isomer of  Co$_{13}$ on graphene.}
\end{figure}

We look at the absolute energy order between the isomer structures on graphene by computing the binding energy $E_{b}$, defined as 
\begin{equation}
\label{bin}
E_{b}=(E_{graphene}+13E_{Co}-E_{Co_{13}/graphene})/13
\end{equation}
where $E_{graphene}$ is the energy of pristine graphene, $E_{Co}$ is the total energy of a single Co atom, and  $E_{Co_{13}/graphene}$ is the energy of the Co$_{13}$ cluster on graphene. The binding energy $E_{b}$ is the energy gained per cobalt atom in the presence of graphene, with respect to a state where they are non bonded among themselves nor to graphene, which is already formed. It allows us to comment on the most stable isomer over graphene.
The binding energy values for the calculated cases are displayed in Fig. \ref{fgr:binding}, a plot that allows us to compare the respective order in energy between all minima of Co$_{13}$ isomers on graphene. 

The highest binding energy corresponds to the icosahedron. Since the icosahedron is not the ground state in free clusters, we found that it is stabilized when interacting with graphene.
Next in energy we find competing and close in energy the $hcp$ $2d$ Co$_{13}$ clusters on graphene. It is worth remembering that the $hcp$ $2d$ isomer is indeed the ground state for the free cluster. The following most stable structures are, in this order, $hcp$ $3d$ and $fcc$; both relaxing to an icosahedral-like geometry. The $pine$ deposited clusters yield the lowest binding energy. In other cases, the binding energies follow the total energy differences between isomers and structural arrangements on graphene. 
It can be seen that the binding energies follow the sequence of isomers already commented in Fig. \ref{fgr:geometries}. The $icosahedron$ and $hcp$ $2d$ cases look related, while all the others follow similar patterns. 

When clusters are deposited on graphene, the $hcp$ $2d$ is no longer the most stable geometry. Now the $icosahedron$ is the most stable one. In fact, we consider that once clusters have been deposited a thermal treatment could somehow drive the clusters from the least stable structures, such as $hcp$ $3d$, $fcc$ and $pine$, to a larger number of the most stable structures, mainly $icosahedra$ and also $hcp$ $2d$. 

We compare the values of the binding energy $E_{b}$ for the most stable geometry over graphene for each isomer with those of the free isomers $E_{b_{free}}$ collected in Table~\ref{tbl:table1}. Following the stability order of the free isomers, namely $hcp$ $2d$, $pine$, $icosahedron$, $fcc$ and $hcp$ $3d$, the values of $E_{b}$ are 3.592, 3.562, 3.602, 3.580 and 3.584 eV, respectively. Because all these $E_{b}$ values are larger than $E_{b_{free}}$ (below 3.51 eV for all the cases) the Co$_{13}$ clusters become more stable when bonded to graphene rather than in the free state.

Now we will focus our study on early stages of Co$_{13}$ cluster deposition by looking at the adsorption energies. The adsorption energies $E_{ad}$ are obtained from the expression
\begin{equation}
\label{ads}
E_{ad}=E_{graphene}+E_{Co_{13}}-E_{Co_{13}/graphene}
\end{equation}
where $E_{Co_{13}}$ is the energy of the isolated cobalt cluster and $E_{graphene}$ and $E_{Co_{13}/graphene}$ are the same as previously defined in equation (\ref{bin}). The adsorption energy represents the process of the bonding between graphene and Co$_{13}$ clusters, the energy released when the cluster binds to graphene. It shows how the cobalt-carbon bonding stabilizes the less stable isomers in the free state. The early state consists of the relaxed free cluster isomers far apart on pristine graphene with almost no interaction between them. These energies suit to the process described in Fig.~\ref{fgr:experimento} at early stages with low coverage. The adsorption energies $E_{ad}$ for the structures of each isomer attached to graphene are plotted in Fig. \ref{fgr:adsorption}. Moving from left to right in Fig. \ref{fgr:adsorption}, the value of $E_{ad}$ increases following the energetic order of the free Co$_{13}$ isomers. Thus a clear correlation is that more stable free isomers have smaller adsorption energies to graphene. Of course, we found over imposed the same energetic trends between structural arrangements of the previously described isomer families. We also get that the $icosahedron$ points shift slightly from a uniform increase. These  $E_{ad}$ trends show that although $hcp$ $2d$ isomer is richer in the Co$_{13}$ cluster beam, its low adsorption energy on graphene implies to stick much less than the icosahedron. These findings add to the highest inherent stability of $icosahedra$ obtained in Fig. \ref{fgr:binding}. It seems that icosahedron isomers are present at large when Co$_{13}$ clusters attach to graphene.

Some authors \cite{Liu2012} propose a correlation between the number of Co-C bonds or the Co-C bond average length and the most stable cluster geometries on graphene. They comment that larger adsorption energies are related to maximizing the number of bonds and minimizing the bond lengths, a shorter bond length means a stronger bond.  In our case, we have that the average Co-C bond lengths for the most stable geometry of each of the five isomers are 2.15, 2.18, 2.19, 2.15 and 2.11 $\mathring{A}$ going from larger to smaller adsorption energy, i.e in the order $hcp$ $3d$, $fcc$, $icosahedron$, $pine$ and $hcp$ $2d$, respectively. We combine these values with the number of bonds for each case: 6, 6, 9, 6 and 3, in the same isomer order given above. Then, we have tried  to explain the order of stability between the different isomers using these ideas. The proposed trends are fulfilled for the three first isomers. However, we find it problematic in the other two cases. The $hcp$ $2d$ has shorter average bond lengths than others, so it should be associated to a larger adsorption energy; it has three bonds with graphene. The higher number of bonds in other cases compensates the larger, and thus fewer stable, average bond lengths. The $pine$ structure has the values of the average bond length and the number of bonds as for $hcp$ $3d$, but the $hcp$ $3d$ has much larger adsorption energy. We thus find that although these trends suggested have been claimed general, they do not always work properly when comparing different isomers.

We have calculated in Fig. \ref{fgr:adsorption} the adsorption energy  of each isomer on graphene for different cluster-graphene arrangements. We are next trying to explain the order of stability between the distinct geometries of each isomer with respect to graphene using the previous ideas. We give the average bond lengths and the number of bonds for all the geometries calculated for the five isomers in Fig. \ref{fgr:geometries}.
For each position of isomers on graphene, it can be seen that in most of the cases the average bond length increase as we move to the right in the row, towards less stable geometries of the isomer over graphene. There are cases where less stable geometries have shorter average bond lengths; in those cases, the number of bonds can help to explain the order of stability. One example is given for the second and third geometries of the $fcc$ isomer; the second geometry becomes more stable due to the large number of bonds, in spite of having larger, at the same time weaker, average Co-C bond length. For $fcc$ and $hcp$ $3d$ cases, the first geometry suffers a structural transformation to an icosahedral-like shape, so it is not reliable to compare them with other geometries of these isomers, which remain like inputs.
Although the combination of the two factors could be considered helpful to explain in some cases the stability order between the different geometries for a certain isomer over graphene, they must not be thought as a general rule because many problematic cases lack explanation as commented in last paragraph.

\subsection{Charge Transfer Between Co$_{13}$ Cluster and Graphene}

We are here looking at the electronic properties related to Co cluster adsorptions. We start by focusing on the charge transfer between Co clusters and graphene. We calculate the local charges in atoms according to Mulliken scheme. Because the same basis sets are used in all cases, we safely comment on the charge differences between them. Figure \ref{fgr:geometries} also includes the electronic charge gained by graphene in all the computed geometries. A positive value indicates excess of electrons, i.e. negatively charged carbon atoms.
In order to discuss the charge transfer in more detail, we are next grouping geometries into two sets: with and without central atom.
For the $icosahedron$, $fcc$ and $hcp$ $3d$ structures, the largest electron transfer is found for the most stable isomer. In fact, starting from one of these three structures the outputs on graphene are icosahedra-like.
For the $pine$ and $hcp$ $2d$ the largest charge transfer is not found for the most stable isomer but for the second most stable. The $pine$ and $hcp$ $2d$ structures have large charge transfers because there are many cobalt atoms bound to graphene, and they have much less spherical shapes. In principle, one can think that the higher the number of cobalt atoms bound to graphene the higher the amount of electrons transferred to graphene, and that the binding between the cluster and graphene is more stable. However, this is not a general rule for the cobalt-graphene binding because we can find larger charge transfers for systems that are not in the ground state. Thus, there must be other important factors that determine which configurations for a group of isomers are the most stable.
Cobalt atoms over graphene prefer to form clusters rather to be spread over graphene because the cobalt-cobalt bonds are stronger than the cobalt-carbon bonds. Summing up, we have two trends fighting against each other. Firstly, the higher number of cobalt atoms bound to graphene the higher the charge transfer, and the higher the stability of the deposited cluster.
Secondly, the higher the number of cobalt atoms in contact with graphene, the lower the strength of the involved cobalt-cobalt bonds because they share charges with carbon atoms. The lower the number of cobalt-cobalt bonds the structure becomes less spherical, so the stability of the systems decreases.
It seems that more detailed analysis should be performed focusing on the free cobalt clusters themselves before they are bound to graphene, in order to elucidate bonding mechanism.

\begin{figure*}[htpb]
\centering
	\includegraphics[width=11.5 cm]{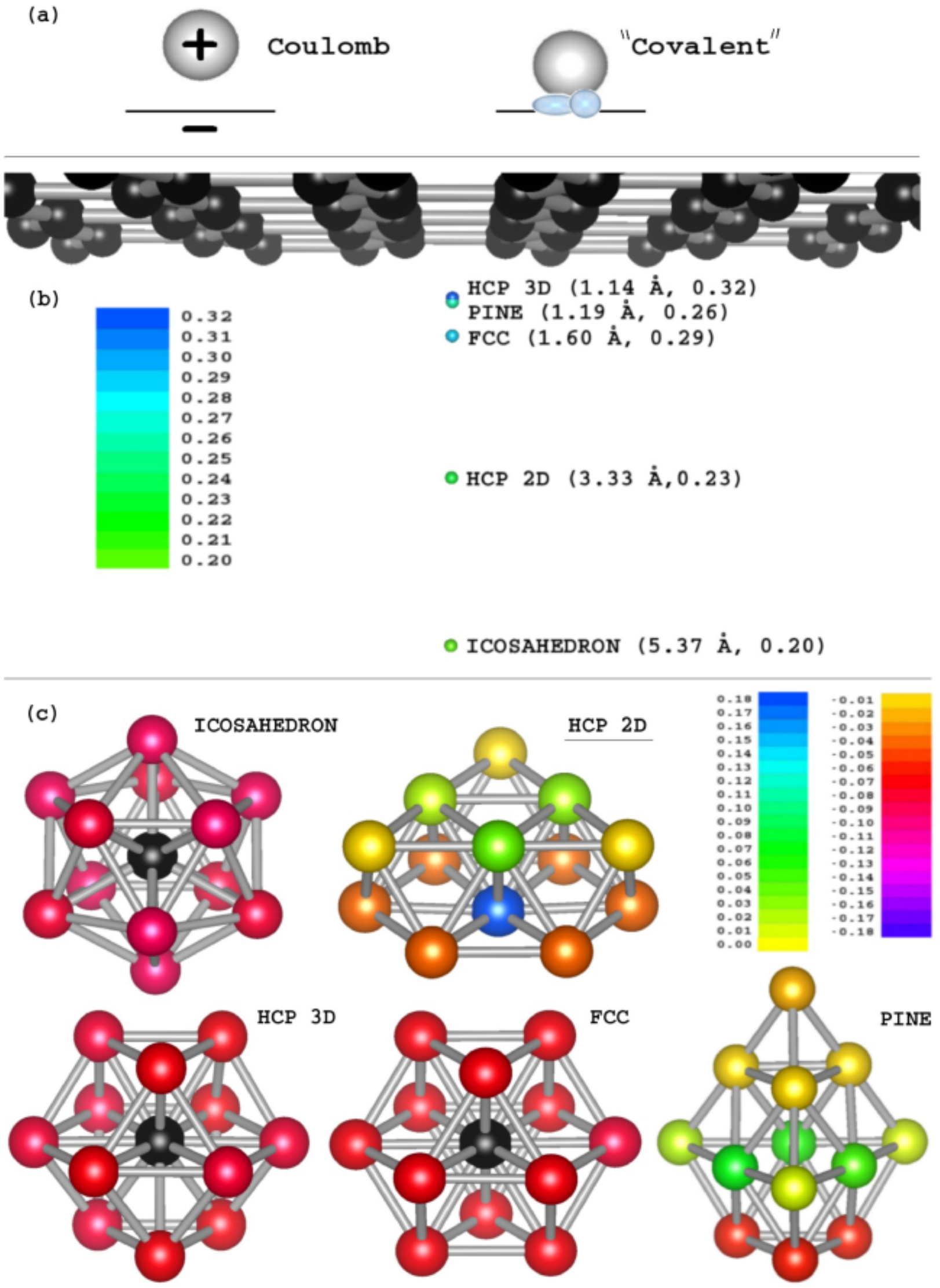}
	\caption{\label{fgr:charge} 
(Colour online) (a) Coulomb and chemical "covalent" binding mechanisms. (b) Scheme of cluster charges as functions of isomers with respect to graphene (c) Local charges on atoms in cluster isomers. In geometries with a central atom, the center atom (in black colour) losses charge to compensate the surface ones. Local charges on the central atom of $icosahedron$, $hcp$ $3d$ and $fcc$ isomers are 1.01, 0.87 and 0.86 e, respectively. 
}
\end{figure*}

\subsection{
Bonding Mechanism}

One of the basic questions we must ask about cluster-graphene charge transfer is: What potential clusters induce in graphene? Of course, we are thinking at the same time about answering when this potential can explain the cluster-graphene bonding. We have started to comment a partial answer to this question in previous paragraphs, where we show that charge transfers affect the order of stability within the structures of a given isomer, but it fails in general to compare arrangements between isomers. It seems that the charge potential affect the stability order. Note that in previous sections, we look at the number of Co-C bonds and state that not always a maximum number of Co-C bonds imply a larger stability.

We shall deal in first order with two main effects: charge transfer and "chemical" bonding. The detailed cluster-graphene bond depends of course, on a correct balance between these effects.
At first, we have looked at the potential induced in graphene by localizing the place where Co cluster charge contributes with zero dipoles, as seen in Fig. \ref{fgr:charge}b; in essence, there, the cluster potential $Q/r$ interact with graphene. The charge from different isomers ranges in the interval 0.2 - 0.3 electrons. The distances, on the other hand, vary more strongly between 1-5 $\AA$. In fact, we have that the potential energy is lower for the more stable $icosahedron$ isomer. Coulomb interactions between clusters and graphene are then not enough to explain the stability of adsorbate Co clusters on graphene.

More important than charges is that there are chemical C-Co bonds between clusters and graphene. We just comment in previous paragraphs that the displaced charge moves to graphene, and in this sense, it acts as an acceptor. However, the Co cluster atoms are not equivalent in all arrangements, because their availability to give charge depends on the strong Co-Co bonds within the cluster.
This trend is measured by looking at the local charges on free isomer clusters, as shown in Fig. \ref{fgr:charge}c. 
The isomers with a central atom are mostly equivalent but with the surface charge increasing in the order of $fcc$, $hcp$ $3d$ and $icosahedron$ isomers. On the other hand, other geometries with all surface atoms have both negative charges in the Co atoms that attract the charge, and positive values in less coordinated atoms depleting the charge.
In the bonding with graphene, the atoms with less charge are prone to form chemical C-Co bonds with graphene.  Thus, not always the number of C-Co bonds is maximized because the charge redistribution taking place in the surface of cobalt isomers.
For chemical C-Co  bonds, it is decisive to consider the charge redistribution within cobalt isomers, much more than the ionic bond based on purely Coulomb interaction of Fig. \ref{fgr:charge}a.

\subsection{
Magnetic Moment}

We have found that when any Co$_{13}$ isomer is deposited on graphene, its magnetic moment decrease. The values of the magnetic moment for the free cluster are included in Table~\ref{tbl:table1} and for the deposited clusters are presented in Fig. \ref{fgr:geometries}. For the $pine$ and $hcp$ $2d$ isomers, the moments decrease an amount of 1.28 and 1.54 $\mu_{B}$, respectively. The decrease for the other three isomers, which end in icosahedra, is even larger.
A general contribution to this decrease for all the isomers comes from electrons transferred to graphene, that induces negative magnetic moment in graphene. However, the magnetic changes are too large to be ascribed to the charge transfer. Charge reorganizes within the cluster by the changes in coordination and bond lengths between clusters and graphene, both determined in the last instance by the geometry.
The magnetic moments for the $pine$ and $hcp$ $2d$ isomers remain almost equal because their equilibrium geometries over graphene keep similar to that of free isomers. 
Since both $hcp$ $3d$ and $fcc$ suffer stronger structural transformations, their magnetic moments change largely. For the $icosahedron$ the magnetic drop is the largest one; in the final disposition over graphene two cobalt atoms move close to graphene, and distort the icosahedron, adapting the values of coordination and bond lengths, which are also important to control the magnetic moment in small clusters.

Cobalt clusters induce magnetic moment in graphene due to the chemical interaction between cobalt and graphene \cite{johll2011graphene}. Figure \ref{fgr:geometries} shows the values of the induced magnetic moment. They vary from -0.06 to -0.58 $\mu_B$ depending mainly on the number of cobalt atoms bonded to graphene. However, other reasons are important such as the geometrical cluster shapes and the cobalt atoms localization over hollow, top or bridge positions. 
The values of 0.5 $\mu_B$ for the induced magnetic moment in graphene are quite large, a finding that could be interesting to develop spintronics devices. We find a direct relationship between the number of cobalt atoms that are bonded to graphene and the induced magnetic moment of graphene: the larger the number of bonds the larger the induced magnetic moment. For four of the five studied isomers, it is possible to reach values for the induced magnetic moment of graphene about -0.5 $\mu_B$. In these arrangements, clusters maximize the number of bonds with graphene. However, these arrangements are not the ground state because as we have already discussed their stability is not directly correlated with the number of bonds.
Anyhow, for the most stable structure ($icosahedron$), the induced magnetic moment of graphene has a remarkable value of -0.33 $\mu_B$, still of interest for applications in spin devices.

\subsection{
DFT+U: on-site interaction}

Here we computed the three most stable configurations of the cluster bonding graphene, $icosahedron$, $hcp$ $2d$ and $hcp$ $3d$, using the DFT+U method. For strongly correlated systems it is advisable to include the on-site interaction to faithfully represent the real system. This is sometimes the case of the d electrons in transition metal atoms such as cobalt.
Our results show that the icosahedron remains as the most stable configuration. Nevertheless, the $hcp3d$ is now more stable than the the $hcp$ $2d$; $hcp$ $3d$ and $hcp$ $2d$ are higher in energy than the icosahedron by 0.15 and 0.82 eV respectively. The structures do not suffer large modifications during relaxation processes, retaining their arrangements.

Concerning the magnetism, the tree systems converge to a total magnetic moment of 26 $\mu_B$, while when not including the on-site interaction they had smaller and different among them values of total magnetic moment. The enhancement of the magnetic moment is expected when the on-site interaction is included.

\subsection{
Bringing results into contact with experiments}

We assess in this section our results against charging effects of both clusters and graphene, as they could be found in experiments.

\subsubsection{
Free Charged Co$_{13}$ Isomers}

In many experimental setups aimed to produce beams of clusters, charged clusters are produced\cite{brucat1986charge, schauer1990production, milani1990improved, scheier1987triply, yoshida1995spin}. In order to extend the validity of our conclusions, we carry out some calculations to determine whether the stability order between free isomers remains the same or not when working with charged clusters.   
To this aim we have computed all the free cobalt isomers both positively and negatively charged, allowing structures to relax. We computed the negative charged clusters to obtain the electron affinity.
We obtained that the $hcp$ $2d$ isomer remains as the most stable one for both the positive and negative charged cluster. The order of stability of the different isomers arranges similarly; except that there is an exchange between $fcc$ and $hcp$ $3d$ for the positive charged cluster and between $pine$ and $icosahedron$ for the negative charged clusters.

These calculations are useful not only to extend the validity of our results according to experiments but also to shed more light over the charge transfer process by computing the work function of the different isomers.
From the total energies of the neutral and charged clusters, we can find the ionization potential and the electron affinity, which allows to obtain the work function as the half of the sum of them. The work function values are 4.33, 4.17, 4.32, 4.35 and 4.21 eV for the $hcp$ $2d$, $pine$, $icosahedron$, $fcc$ and $hcp$ $3d$ isomers, respectively.
There are no large differences between the work function of the different isomers, which agrees with the fact that the charge transfer values between clusters and graphene do not vary a lot between isomers.
This result agrees with the idea that the chemical bonds between cobalt and carbon atoms determine specifically the most stable bonding arrangement, more than the type of isomer. Indeed, the stability of the composites is not dominated by the Co-Co interaction but by the Co-C carbon interaction. In this work, we show that placing the $hcp$ $2d$ isomer over one face or another is not equivalent at all. Note that having a large number of bonds does not guarantee to the cluster disposition on graphene being more stable. The charge state of the cobalt atoms in the free isomers appears to be preponderant concerning the cobalt carbon bonding. So in the end, the cobalt carbon bonding plays a key role. Furthermore, if the Co-Co interaction were the predominant part, the $hcp$ $2d$ isomer should be the most stable structure over graphene (as it is the most stable in the free state), and it is not the case.

\subsubsection{
Doped Graphene}

In experiments, graphene is usually found over a substrate, which we have not taken into account in our calculations. The main effect of this substrate is to raise or lower the Fermi level by doping graphene. In order to check the effect a substrate could have on our previously reported conclusions, we have to study the doping effect of graphene by a supporting substrate on the stability order for the different isomers.
We have added or removed a charge from the graphene-isomer system either globally or locally following two different schemes. Firstly, we specify one more (or less) charge directly in the program input to modify the amount of net charge. Secondly, we substitute one carbon atom in the unit cell by a neighbor atom in the periodic table. We dope by a boron atom so the system gets positively charged, and by a nitrogen atom to get it negatively charged. Because we here study the bonding between the cluster and graphene, we have to be careful to avoid direct interaction between the cobalt clusters and boron and nitrogen atoms. For this reason, we replace a carbon atom as far as possible from the cluster within the unit cell area.
We have carried out these calculations with three distinct isomers that we got on graphene in the minimum geometry: $icosahedron$, $hcp$ $2d$ and $pine$. Note that both $fcc$ and $hcp$ $3d$ structures on graphene relax into icosahedra-like geometries.

\begin{figure}[h]
\centering
	\includegraphics[width=8cm]{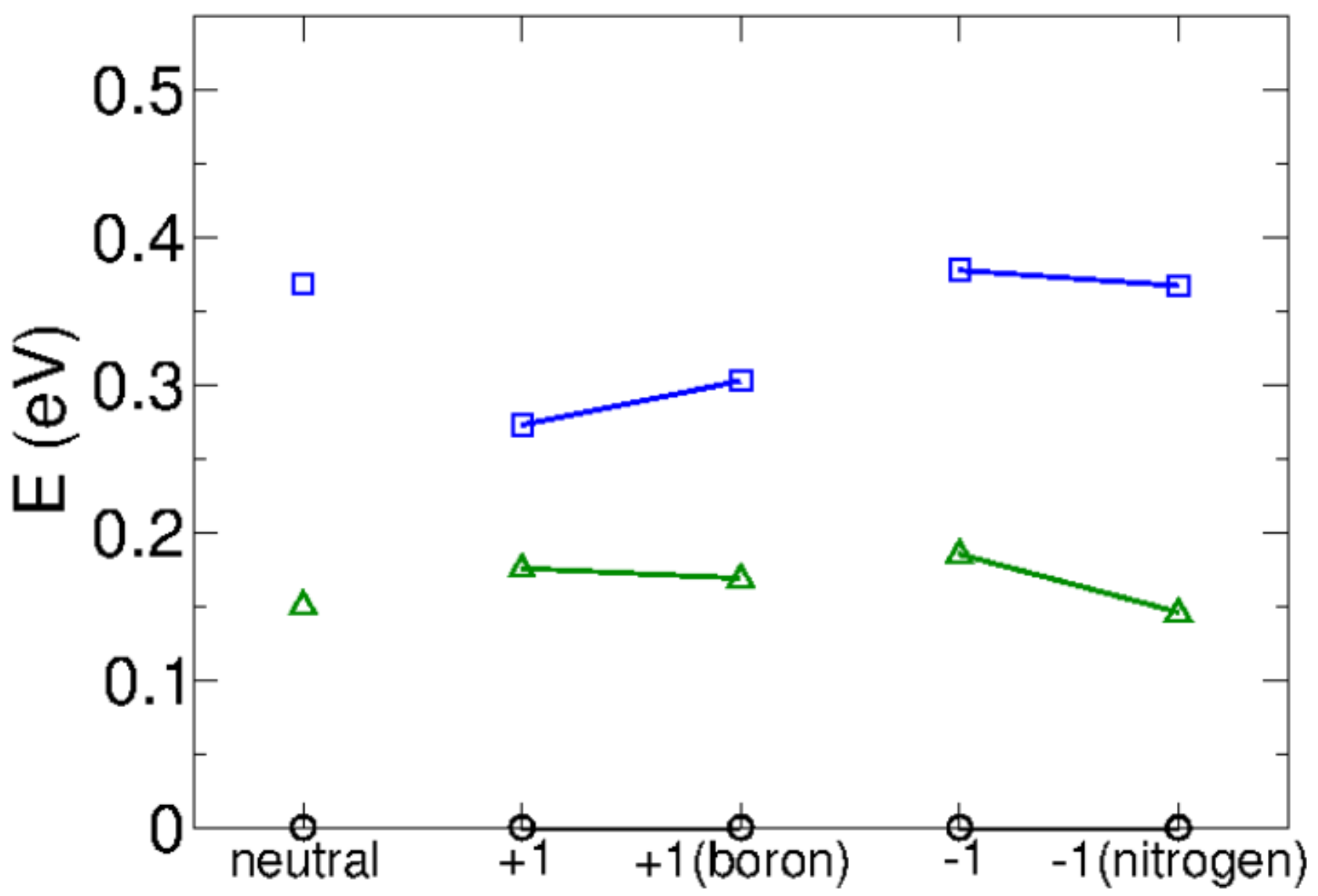}
	\caption{\label{fgr:sustrato} 
(Colour online) Differences in energy between the $icosahedron$ (circles), $hcp$ $2d$ (triangles) and $pine$ (squares) isomers on graphene for different charge states by changing uniformly the Fermi level or by doping with atoms.}
	\end{figure}

Figure \ref{fgr:sustrato} shows the order of stability when the graphene-cluster system is charged and the differences in energy between the isomers over graphene. 
Note that we can compare these energy differences between isomers over graphene with those values obtained for the non-charged system.
For the case of the positively charged system in either of the two used schemes, we found that the energy difference between the $icosahedron$ and the $hcp$ $2d$ becomes larger while for the $hcp$ $2d$ and the $pine$ it becomes smaller; the difference between the $icosahedron$ and the $pine$ also becomes smaller. 
For the negatively charged system, all the energy differences become larger for the first method used, while they stay approximately the same for the second method.
Anyhow, the relative order of stability between the distinct isomers is preserved in all the cases, with the $icosahedron$ as the most stable structure followed by the $hcp$ $2d$ and the $pine$ ones.

\section*{Conclusions}

We have calculated several relative positions of each low-lying Co$_{13}$ isomer on graphene. We found that the Co$_{13}$ clusters bonded to graphene becomes the most stable isomer as $icosahedra$. Note that the free-standing ground-state isomer for the Co$_{13}$ clusters is a bi-planar structure, we have named $hcp$ $2d$, as it was discovered recently \cite{faustino-star_like};  apart from $hcp$ $2d$, also the $pine$ isomer resulted more stable than the $icosahedron$ isomer when they are unsupported, which used to be normally proposed as the most stable isomer for free standing calculations.
We have next analysed the charge character of the atoms in the free isomers for all those geometries over graphene. We conclude that isomers prefer to bond graphene by the cobalt atoms that prefer to give charge. In all the cases studied, there is charge transfer from the cluster towards graphene.
However, we have checked that this charge transfer is not the key mechanism for the cluster-graphene bonding, which instead of ionic is strongly chemical.
In principle, one could guess that the findings would be applicable for different size clusters. However, since we are working at the nanoscale, there could be some sizes where other factors destroy this cobalt-carbon rule due to the important dependence on the size. This appears as an interesting new study, but the evolution of the trends we have found in the size would represent a new project, and it is out of the scope of the present work.
When the clusters bond to graphene, it becomes magnetic in all the cases, with relatively high induced magnetic moments of about -0.3 $\mu_B$ for the most stable structure.
Note that free charged clusters both positively and negatively charged, have no drastic changes in the isomer stability, with the $hcp$ $2d$ structure remaining as the ground state.
In order to discuss more realistic experimental setups, we have last studied the effect of a substrate below graphene by raising and lowering the Fermi level and by doping graphene with neighbor atoms. Our results let us to affirm that the stability order among isomers does not change essentially under both kinds of doping, a finding that gives even wider validity to our conclusions.

\section*{Computational details}

We use the SIESTA (Spanish Initiative for Electronic Simulations with Thousands of Atoms) code to carry out density functional theory (DFT) calculations of Co$_{13}$ clusters free and supported on graphene. For the exchange and correlation potentials we used the Perdew-Burke-Ernzenhof form of the generalized gradient approximation (GGA).\cite{perdew1996generalized} Firstly, we relax graphene in the unit cell that we are using to deposit Co$_{13}$ clusters. Not only the atomic positions but the unit cell itself is also allowed to relax. From the analysis in previous test calculations \cite{Liu2012}, we know that the unit cell has to be large enough in order to avoid cluster-cluster interaction. We used a coverage of 21.6\%,  with the coverage defined as $(N_{Co}/N_{C})\times100$, where $N_{Co}$ and $N_{C}$ are the numbers of the cobalt and carbon atoms, respectively. We start picking a bidimensional unit cell of $12.96\mathring{A}\times12.47\mathring{A}$ separated by an empty space between supercell replica converged as non interacting to a distance of 16 $\mathring{A}$. 
Secondly, the geometries of the clusters were relaxed together with the unit cell.
We use the same computational parameters in all the calculations: an electronic temperature of 25 meV, a meshcutoff of 250 Ry and an energy shift of 10 meV. We use the Monkhorst-Pack method for the choice of the k-point mesh: 2x2x1. 
The atomic cores were described by nonlocal norm-conserving relativistic Troullier-Martins\cite{troullier1991efficient} pseudopotentials factorized in the Kleynman-Bylander form.
For the cobalt atom we use radii r$_{c}$=1.06 $\mathring{A}$; for the carbon atom, we use radii r$_{s}$=0.79, r$_{p}$=0.79 and r$_{d}$=1.16 $\mathring{A}$. The pseudopotential for cobalt was generated using the valence configuration 4s$^{2}$3d$^{7}$. We have included non-linear core corrections with a matching radius of 0.75 a.u..
The pseudopotentials have been tested so that they reproduce accurately the eigenvalues of different excited states of the isolate atom. The cobalt pseudopotential is also tested by computing the bulk; we obtained that the $hcp$ structure is preferred over the $fcc$ by 0.025eV/atom. The calculated $hcp$ structure has a first neighbors distance of 2.54 $\mathring{A}$ and a magnetic moment per atom of 1.64$\mu_B$, which are in agreement with experimental results.\cite{myers1951spontaneous}
The basis size is double zeta plus polarization orbitals. We use a basis whose cutoff radii values in $\mathring{A}$ for the cobalt atom are 4.23 and 3.47 in the 4s orbital, 4.23 in the 4p orbital and 4.23 and 1.28 in the 3d orbital. For the carbon atom, basis cutoff radii in $\mathring{A}$ are 3.15 and 1.86 in the 2s orbital and 4.04 and 2.06 in the 2p orbital.

When we deposit the cluster relaxed geometry on the unit cell of graphene we relax both the inner atoms and the unit cell. For each Co$_{13}$ isomer interacting with graphene, different geometry arrangements are possible. Based on coordination, we start by matching atoms, edges or faces with one or few carbon atoms, respectively. Depending on symmetry, free cluster atoms are bonded to graphene in several arrangements e.g.with a metal atom at the center of hexagonal faces or edges. Thus, each isomer relaxes to the closest lying minima on graphene in many arrangements until forces are below 0.006 eV/$\mathring{A}$.

In order to check the influence of on-site interaction on the systems under study here we carried out DFT+U calculations using the VASP code. We computed some of the configurations again starting from the relaxed geometries obtained with SIESTA and allowing them to relax including a U-J difference of 3 eV.

\begin{acknowledgments}
This work has been partially supported by the Project FIS2013-48286-C2-1-P of the Spanish Ministry of Economy and Competitiveness MINECO, 
 the Basque Government under the ETORTEK Program 2014 (nanoGUNE2014), and the University of the Basque Country (Grant No. IT-366-07). 
TAL acknowledges the grant of the MPC Material Physics Center- San Sebasti\'a n. 
FAG acknowledge the DIPC for their generous hospitality and the kindly help on the computational assistance of Diego Lasa. We also want to acknowledge the DIPC computer center (TAL, AAF and FAG).

\end{acknowledgments}

%

\end{document}